\documentclass[aps,prd,reprint]{revtex4-2}

\usepackage{graphicx}
\usepackage{dcolumn}
\usepackage{bm}
\usepackage{amssymb}
\usepackage{natbib}

\begin{document}

\title{Strong-coupling of Minkowski spacetime in $f(R)$ gravity: a Hamiltonian perspective}

\author{Aldair Pantoja}
\email[]{jpantoja@ifuap.buap.mx}
\affiliation{Instituto de F\'isica, Benem\'erita Universidad Aut\'onoma de Puebla. \\ Apartado Postal J-48 72570, Puebla Pue., M\'exico.}

\date{\today}

\begin{abstract}
A Hamiltonian analysis of $f(R)$ gravity in the perturbation field approximation is performed. The action principle is presented as a series expansion around a generic fixed background with constant scalar curvature $R_{0}$, then we restrict the theory to maximally symmetric spaces, later we recast the action by selecting $R_{0}=0$, i.e., by choosing Minkowski as the background. The Hamiltonian formulation is constructed by following the appropriate methods to approach higher-order systems. The constraint structure is identified for several cases that arise depending on imposed conditions on the values of the function $f(R)$ and its derivatives, classification of constraints allows to make a proper count of degrees of freedom for each case, at the end we confirm the spectrum behaviour of $f(R)$ gravity around Minkowski spacetime that has been reported by another means in the recent literature.
\end{abstract}

\maketitle

\section{Introduction}

Since the birth of Einstein's general theory of relativity \cite{Einstein1,Einstein2}, a large number of modifications of the theory have been proposed, those modifications comes from a diverse kind of reasons, such as: to render the theory as a renormalizable one in the pursuit of a quantum theory of gravity \cite{Stelle1}, to explain the distribution of matter and energy in the universe \cite{Koyama1}, to clarify the accelerated expansion of the universe \cite{Koyama2}, to remove spacetime singularities \cite{Bosma,Allaithy}, etc. It is because of this diversity of motivations that we have so many faces of the so-called modified gravity, for example: conformal gravity as an attempt to unify gravity with electromagnetism \cite{Weyl}, topological massive gravity where the Chern-Simons invariant is coupled to the Einstein-Hilbert action \cite{Deser}, Ho\v{r}ava's gravity where four dimensional diffeomorphisms are abandoned \cite{Horava}, etc. Experience taught us that for the \textit{problems} of General Relativity that modified gravity tries to resolve, many others problems arise in most cases if not in all. Despite its comings and goings, it should be remembered that a modified theory of gravity have its importance if not as an ultimately theory, but as point of comparison and as platform to explore novel ideas or phenomena that General Relativity alone cannot display. In this regard, recent works have shown that the so-called $f(R)$ gravity can admit strongly-coupled backgrounds, a feature that is not present in General Relativity, lets elaborate more on this issue.

The $f(R)$ gravity models are theories where the conventional action $\int R\sqrt{-g}d^{4}x$ is replaced with $\int f(R)\sqrt{-g}d^{4}x$, being $f(R)$ a function of the scalar curvature $R$. Many aspects of $f(R)$ gravity has been explored over the years since it was first introduced by Buchdahl \cite{Buchdahl} to highlight some cosmological aspects of it, a comprehensive guide to $f(R)$ gravity can be found in \cite{Sotiriou1} and in the references therein. It is a well-known result that generic $f(R)$ theories propagate 3 degrees of freedom associated to two polarizations of massless spin-2 modes and to the so-called scalaron mode but, it has been showed recently that the particular case $f(R)=\alpha R^{2}$ ($\alpha$ a constant), presents a kind of discontinuity in the number of degrees of freedom depending upon how one approaches the model, namely, if one tries to read the spectrum by taking the full theory and also by taking the weak field approximation around Minkowski space then a discrepancy arises: 3 degrees of freedom for full theory and no propagating modes for perturbation theory (even if one tries to recover the full theory by considering higher-orders of the perturbation) \cite{Hell,Barker}. These findings suggests that there exists some $f(R)$ models where, if some perturbation field approximation is made, then the resulting degrees of freedom propagating on top of the background can depend on what background is considered, or by recalling the terminology given in \cite{Delhom}, the background can exhibit a strong-coupling instability, or equivalently, the background can be strongly-coupled. It turns out that the preceding suggestion is true, $f(R)=\alpha R^{2}$ is only a particular choice of a more broad range of $f(R)$ models that are compatible with a strongly-coupled Minkowski spacetime.

In \cite{Casado1,Casado2}, the subject is analysed in the context of constant-curvature solutions of $f(R)$ gravity, a pertinent distinction between various $f(R)$ models is introduced in order to state the main results, namely, for a certain constant-curvature scalar $R_{0}$, if the function $f(R)$ satisfies $f'(R_{0})=0$ (and also $f(R_{0})=0$ as a consequence of the equations of motion) then the $f(R)$ model is called $R_{0}-$\textit{degenerate}, on the contrary, if $f'(R_{0})\neq0$ then it is called $R_{0}-$\textit{non-degenerate}. Then, it is demonstrated that in the $(R_{0}=0)-$\textit{degenerate} case the Minkowski spacetime is strongly-coupled, i.e. the $f(R)$ model lacks of propagating degrees of freedom in the perturbation field approximation, regardless of whether $f''(0)=0$ or $f''(0)\neq0$. Similarly, it is claimed that the $R_{0}-$\textit{non-degenerate} models propagates the usual 3 degrees of freedom provided that $f''(R_{0})\neq0$ and only 2 degrees of freedom if $f''(R_{0})=0$, the later results seem to be independent whether $R_{0}=0$ or not.

All of these outcomes represent a motivation to pursue a Hamiltonian analysis in order to study the spectrum of different kinds of $f(R)$ models, in particular, of the $(R_{0}=0)-$\textit{non-degenerate} and $(R_{0}=0)-$\textit{degenerate} cases. It is well established that the Hamiltonian formalism represents a good tool to extract the number of physical degrees of freedom of a constrained theory \cite{Dirac1} and, since $f(R)$ gravity is a constrained one then it is natural to ask for the Hamiltonian aspects of it. It is worth mentioning that in fact, several Hamiltonian approaches of $f(R)$ gravity are present in the literature \cite{Deruelle,Liang} but, the issue on the discontinuity on the degrees of freedom is not addressed. Given this line of though, the object of this work is to develop a constraint analysis of the $(R_{0}=0)-$\textit{non-degenerate} and $(R_{0}=0)-$\textit{degenerate} $f(R)$ models with the aim on to identify their spectrum, and verify that:
\begin{enumerate}
\item{$(R_{0}=0)-$\textit{non-degenerate} models propagate 3 degrees of freedom as long as $f''(0)\neq0$.}
\item{$(R_{0}=0)-$\textit{non-degenerate} models propagate 2 degrees of freedom when $f''(0)=0$.}
\item{$(R_{0}=0)-$\textit{degenerate} models lacks of physical degrees of freedom regardless of whether $f''(0)=0$ or $f''(0)\neq0$.}
\end{enumerate}

The paper is organized as follows: in Section II we establish the action principle by assuming that the function $f(R)$ admits a series expansion around some value $R=R_{0}$, then we perform the perturbation field approximation and we restrict the background to a maximally symmetric one, in Section III we develop the Hamiltonian formalism of the model around $R_{0}=0$, i.e. around Minkowski spacetime, then we obtain and classify all the constraints of the theory, this classification allow us to perform the counting of degrees of freedom for particular cases depending upon the values of $f'(0)$ and $f''(0)$, in Section IV final discussion and comments are made.

\section{The action}

Lets begin with the standard form of the action principle for $f(R)$ gravity in the metric formulation in four dimensions
\begin{equation}
\label{Action}
S = \frac{1}{2\kappa}\int f(R)\sqrt{-g}d^{4}x
\end{equation}
where $\kappa=8\pi G$ being $G$ the gravitational constant, $f(R)$ a generic function of the scalar curvature $R$ and $g$ the determinant of the metric $g_{\mu\nu}$. Variation of action (\ref{Action}) with respect to the metric produce the equations of motion
\begin{equation}
f'(R)R_{\mu\nu} - \frac{f(R)}{2}g_{\mu\nu} - (D_{\mu}D_{\nu} - g_{\mu\nu}\square)f'(R) = 0
\end{equation}
where $f'(R)$ denotes the derivative of the function $f$ with respect to $R$, $D_{\mu}$ is the covariant derivative with respect to metric $g_{\mu\nu}$. Substitution of $R=R_{0}$ in the contracted equations gives the relation
\begin{equation}
\label{Condition}
2f(R_{0}) = f'(R_{0})R_{0}
\end{equation}
which serve us to classify the $f(R)$ models, as presented in \cite{Casado1,Casado2}. The $f(R)$ theories that satisfy the conditions
\begin{eqnarray}
f(R_{0}) &=& 0,
\\
f'(R_{0}) &=& 0,
\end{eqnarray}
are called $R_{0}-$\textit{degenerate}, on the other hand, if $f'(R_{0})\neq0$ then it is called $R_{0}-$\textit{non-degenerate}. One can make another distinction by considering the particular choice $R_{0}=0$, this give rise to $(R_{0}\neq0)-$(\textit{non-})\textit{degenerate} and $(R_{0}=0)-$(\textit{non-})\textit{degenerate} models, for the Hamiltonian analysis presented in the next section we will focus on the later kind of models.

Now lets express the action as a series expansion around a background of metric $g_{\mu\nu}^{(0)}$, more precisely
\begin{equation}
\label{Expansion}
f(R)=f(R_{0})+f'(R_{0})(R-R_{0})+\frac{1}{2}f''(R_{0})(R-R_{0})^{2}+\cdots
\end{equation}
then, by inserting equation (\ref{Expansion}) into the action (\ref{Action}) we get
\begin{eqnarray}
\nonumber
S &=& \int\Big(\big[f(R_{0}) - f'(R_{0})R_{0} + \frac{1}{2}f''(R_{0})R_{0}^{2}\big] + \big[f'(R_{0})
\\
\label{ActionExpansion}
&& -f''(R_{0})R_{0}\big]R + \frac{1}{2}f''(R_{0})R^{2} \Big)\sqrt{-g}d^{4}x
\end{eqnarray}
where we omitted the global factor $(2\kappa)^{-1}$ and only considered the expansion up to quadratic terms in $R$. Now we need to express $R$ and $R^{2}$ in a suitable fashion to make the Hamiltonian analysis, for such purpose we consider small perturbations around the fixed background
\begin{equation}
g_{\mu\nu} = g_{\mu\nu}^{(0)} + h_{\mu\nu}
\end{equation}
with inverse metric given by $g^{\mu\nu}=g^{\mu\nu}_{(0)} - h^{\mu\nu} + h^{\mu\rho}h_{\rho}{}^{\nu}$ and, from now and on every quantity indicated with label ${}^{(0)}$ or ${}_{(0)}$ will indicate a quantity with respect to the background. We can calculate the scalar $R$ from the following expansions of the Riemman and full covariant Riemman tensors
\begin{eqnarray}
R^{\rho}{}_{\mu\sigma\nu} &=& R^{\rho}_{(0)\mu\sigma\nu} + \delta R^{\rho}{}_{\mu\sigma\nu},
\\
R_{\rho\mu\sigma\nu} &=& R_{\rho\mu\sigma\nu}^{(0)} + \delta R_{\rho\mu\sigma\nu}
\end{eqnarray}
where $\delta R_{\rho\mu\sigma\nu} = h_{\rho\alpha}R^{\alpha}_{(0)\mu\sigma\nu} + g_{\rho\alpha}^{(0)}\delta R^{\alpha}{}_{\mu\sigma\nu}$. The variation of the Riemann tensor $\delta R^{\rho}{}_{\mu\sigma\nu}$ is straightforward if we take in account that $\delta g_{\mu\nu}=h_{\mu\nu}$. The contraction $g^{\mu\nu}R_{\rho\mu\alpha\nu}$ produce the Ricci tensor, another contraction with $g^{\rho\alpha}$ gives the scalar curvature:
\begin{eqnarray}
R_{\mu\nu} &=& R^{(0)}_{\mu\nu} + \nabla_{\sigma}\nabla_{(\nu}h^{\sigma}{}_{\mu)} - \frac{1}{2}(\square h_{\mu\nu} + \nabla_{\mu}\nabla_{\nu}h),
\\
\nonumber
R &=& R_{0} - h^{\mu\nu}R_{\mu\nu}^{(0)} + h^{\mu\rho}h_{\rho}{}^{\nu}R^{(0)}_{\mu\nu} + \nabla_{\mu}\nabla_{\nu}h^{\mu\nu} - \square h
\\
\nonumber
&& + \frac{1}{2}h^{\mu\nu}\square h_{\mu\nu} + \frac{1}{2}h^{\mu\nu}\nabla_{\mu}\nabla_{\nu}h - h^{\mu\nu}\nabla_{\rho}\nabla_{\mu}h^{\rho}{}_{\nu},
\\
\label{R}
\end{eqnarray}
where $R_{\mu\nu}^{(0)}$ stands for $g_{(0)}^{\rho\sigma}R_{\rho\mu\sigma\nu}^{(0)}$, $\nabla_{\mu}$ is the covariant derivative with respect to the metric $g_{\mu\nu}^{(0)}$, the $\square$ operator is defined via $\square=g_{(0)}^{\mu\nu}\nabla_{\mu}\nabla_{\nu}$, the trace of the perturbation field is $h=h^{\mu}{}_{\mu}$, $R^{(0)}=R_{0}$ and we retain up to second-order terms in the perturbation field.

Now, as for the quadratic term, a similar procedure (or by squaring (\ref{R}) and reordering properly) lead us to the following expansion
\begin{equation}
\label{R^2Original}
R^{2} = (\nabla_{\mu}\nabla_{\nu}h^{\mu\nu} - \square h + \cdots)^{2} + 2R_{0}(R-R_{0}) + R_{0}^{2}
\end{equation}
where $R$ is given by the expansion (\ref{R}). We can introduce an auxiliary field $\phi$ in order to recast the first term of (\ref{R^2Original}) as
\begin{equation}
\label{R^2}
2\big[ (\nabla_{\mu}\nabla_{\nu}h^{\mu\nu} - \square h - \cdots)\phi - \frac{1}{2}\phi^{2} \big],
\end{equation}
in this reformulation the equations of motion of $\phi$ produce the expression $\phi=\nabla_{\mu}\nabla_{\nu}h^{\mu\nu} - \square h - \cdots$ so, after substitution of this relation in the action one recover its original fashion. It is important to comment that the field $\phi$ is introduced here as dynamical field and not as a constraint enforced by some Lagrange multiplier, this distinction is important to make a correct counting of degrees of freedom, we can anticipate of course that $\phi$ and its momentum will contribute to the constraints because of this dynamical footing.

Up to this point, the expressions detailed in (\ref{R}) and (\ref{R^2Original}) are valid for every background with constant $R_{0}$, now we will restrict to a particular case: maximally symmetric spacetimes. Maximally symmetric spacetimes are those spaces that possesses the largest possible number of symmetries, in these type of spaces the scalar curvature is constant and the Riemman and Ricci tensors are directly given by a combination of the metric tensor
\begin{eqnarray}
\label{MS1}
R_{\rho\mu\sigma\nu}^{(0)} &=& \frac{R_{0}}{12}\big(g_{\rho\sigma}^{(0)}g_{\mu\nu}^{(0)} - g_{\rho\nu}^{(0)}g_{\mu\sigma}^{(0)}\big),
\\
\label{MS2}
R_{\mu\nu}^{(0)} &=& \frac{R_{0}
}{4}g_{\mu\nu}^{(0)}.
\end{eqnarray}
Then, if we substitute the expressions (\ref{MS1}) and (\ref{MS2}) in both $R\sqrt{-g}$ and $R^{2}\sqrt{-g}$ and by taking the expansion $\sqrt{-g} = \big( 1 + \frac{1}{2}h + \frac{1}{8}h^{2} - \frac{1}{4}h^{\mu\nu}h_{\mu\nu} + \cdots\big)\sqrt{-g^{(0)}}$ we obtain the following form for the action
\begin{widetext}
\begin{eqnarray}
\nonumber
S &=& \int\Big( \big[f(R_{0}) - f'(R_{0})R_{0} \big]\big[1+\frac{1}{2}h+\frac{1}{4}(\frac{1}{2}h^{2}-h^{\mu\nu}h_{\mu\nu})\big]
\\
\nonumber
&& + f'(R_{0})\big[R_{0} + \frac{R_{0}}{4}h + \frac{1}{2}h^{\mu\nu}\square h_{\mu\nu} - h^{\mu\nu}\nabla_{\rho}\nabla_{\mu}h^{\rho}{}_{\nu} + h\nabla_{\mu}\nabla_{\nu}h^{\mu\nu} - \frac{1}{2}h\square h \big]
\\
\label{Action2}
&& + f''(R_{0})\big[(\nabla_{\mu}\nabla_{\nu}h^{\mu\nu} - \square h - \frac{1}{4}R_{0}h)\phi - \frac{1}{2}\phi^{2}\big] \Big)\sqrt{-g_{(0)}}d^{4}x.
\end{eqnarray}
\end{widetext}
If we vary the action (\ref{Action2}) with respect to the perturbation field $h_{\mu\nu}$, then we obtain the following equations of motion:
\begin{equation}
\label{LEoM}
f'(R_{0})\phi_{\mu\nu} + f''(R_{0})(\nabla_{\mu}\nabla_{\nu} - g_{\mu\nu}^{(0)}\square - \frac{1}{4}R_{0}g_{\mu\nu}^{(0)})\phi = 0,
\end{equation}
where $\phi_{\mu\nu}$ is defined by
\begin{eqnarray}
\nonumber
\phi_{\mu\nu} &=& \frac{1}{2}\big[\square h_{\mu\nu} + \nabla_{\mu}\nabla_{\nu}h - (\nabla_{\rho}\nabla_{\mu}h^{\rho}{}_{\nu}+\nabla_{\rho}\nabla_{\nu}h^{\rho}{}_{\mu})
\\
\nonumber
&& - g_{\mu\nu}^{(0)}(\square h - \nabla_{\alpha}\nabla_{\beta}h^{\alpha\beta})\big] - \frac{R_{0}}{4}(\frac{1}{2}g_{\mu\nu}^{(0)}h - h_{\mu\nu}).
\\
\label{EinsteinTensor}
\end{eqnarray}
It is worth commenting that there are hidden terms $\frac{R_{0}}{12}h^{2}-\frac{R_{0}}{3}h^{\mu\nu}h_{\mu\nu}$ in (\ref{Action2}) and the subsequent terms $\frac{R_{0}}{12}g_{\mu\nu}^{(0)}h-\frac{R_{0}}{3}h_{\mu\nu}$ in (\ref{EinsteinTensor}) due to the double covariant derivative $-\nabla_{\rho}\nabla_{\mu}h^{\rho}{}_{\nu}$, if these terms are not present is just a matter of presentation. The equations of motion (\ref{LEoM}) are exactly the same as those presented in \cite{Casado2}, thus, all results therein derived from the linearised equations of motion also follows from (\ref{Action2}), in fact, if we take the trace of equation (\ref{LEoM}) we obtain
\begin{equation}
\label{TLEoM}
f''(R_{0})(\square + \frac{R_{0}}{3})\phi - \frac{1}{3}f'(R_{0})\phi = 0,
\end{equation}
in accordance to \cite{Casado2}. In the next section we will take the action (\ref{Action2}) to develop the Hamiltonian formulation to extract the number of degrees of freedom of the $(R_{0}=0)-$\textit{non-degenerate} and $(R_{0}=0)-$\textit{degenerate} $f(R)$ models.

\section{Hamiltonian formulation}

We begin this section by setting $R_{0}=0$, the only possible background for this value is the Minkowski spacetime, then $g_{\mu\nu}^{(0)}\rightarrow\eta_{\mu\nu}$, $\nabla_{\mu}\rightarrow\partial_{\mu}$ and $\sqrt{-g_{(0)}}\rightarrow 1$. The action recasts as follows
\begin{eqnarray}
\label{LAction}
S &=& \int\Big( f(0)\big[1+\frac{1}{2}h+\frac{1}{4}(\frac{1}{2}h^{2}-h^{\mu\nu}h_{\mu\nu})\big]
\\
\nonumber
&& + f'(0)\big[ \frac{1}{2}h^{\mu\nu}\square h_{\mu\nu} - h^{\mu\nu}\partial_{\rho}\partial_{\mu}h^{\rho}{}_{\nu} + h\partial_{\mu}\partial_{\nu}h^{\mu\nu}
\\
\nonumber
&&  - \frac{1}{2}h\square h \big] + f''(0)\big[(\partial_{\mu}\partial_{\nu}h^{\mu\nu} - \square h)\phi - \frac{1}{2}\phi^{2}\big] \Big)d^{4}x.
\end{eqnarray}
From variation of the action (\ref{LAction}) (or by virtue of (\ref{TLEoM})) we get a Klein-Gordon-type equation
\begin{equation}
\label{KG}
(\square - m^{2})\phi = 0
\end{equation}
where $m^{2}\equiv\frac{f'(0)}{3f''(0)}$, if $f'(0)=0$ and $f''(0)\neq0$ then (\ref{KG}) entails a massless scalar mode.

The presence of the massless scalar mode is only apparent, as noted in \citep{Casado2}, we will reach to the same conclusion from the Hamiltonian analysis.

Now, with the aim to define the canonical momenta we need to perform a $3+1$ decomposition, by adopting a signature $(-+++)$ the corresponding splitting gives
\begin{widetext}
\begin{eqnarray}
\nonumber
S &=& \int \Big[ f(0)\Big( 1 - \frac{1}{2}h_{00} + \frac{1}{2}h^{i}{}_{i} - \frac{1}{8}(h_{00})^{2} - \frac{1}{4}h_{00}h^{i}{}_{i} + \frac{1}{8}(h^{i}{}_{i})^{2} - \frac{1}{2}h^{0i}h_{0i} - \frac{1}{4}h^{ij}h_{ij} \Big)
\\
\nonumber
&& + f'(0)\Big( \frac{1}{2}\dot{h}^{ij}\dot{h}_{ij} - \frac{1}{2}\dot{h}^{i}{}_{i}\dot{h}^{j}{}_{j} + h^{0i}\nabla^{2}h_{0i} + \frac{1}{2}h^{ij}\nabla^{2}h_{ij} + 2\dot{h}^{i}{}_{i}\partial^{j}h_{0j} - h_{00}\partial_{i}\partial_{j}h^{ij}
\\
\nonumber
&& + h^{k}{}_{k}\partial_{i}\partial_{j}h^{ij} - 2\dot{h}^{ij}\partial_{i}\partial_{i}h_{0j} + h_{0i}\partial^{i}\partial^{j}h_{0j} - h_{i}{}^{k}\partial^{i}\partial_{j}h^{j}{}_{k} + h_{00}\nabla^{2}h^{i}{}_{i} - \frac{1}{2}h^{i}{}_{i}\nabla^{2}h^{j}{}_{j} \Big)
\\
&& + f''(0)\Big( \big[\ddot{h}^{i}{}_{i} - 2\partial^{i}\dot{h}_{0i} + \partial^{i}\partial^{j}h_{ij} + \nabla^{2}h_{00} - \nabla^{2}h^{i}{}_{i}\big]\phi - \frac{1}{2}\phi^{2} \Big) \Big]d^{4}x,
\end{eqnarray}
\end{widetext}
where $\nabla^{2}=\partial^{i}\partial_{i}$ and $\dot{h}_{\mu\nu}=\partial_{0}h_{\mu\nu}$ stands for derivative of the perturbation field with respect to $t$.

We can immediately note the expected higher-order behaviour in the first term associated to $f''(0)$, this detail demands the appropriate tools to formulae a Hamiltonian theory, we will follow Gitman, Lyakovich and Tyutin \cite{Gitman}. First, we introduce the extra fields $s_{\mu\nu}$ and $v_{\mu\nu}$ via the following definitions
\begin{eqnarray}
\label{GLT1}
s_{\mu\nu} &\equiv& \dot{h}_{\mu\nu},
\\
\label{GLT2}
v_{\mu\nu} &\equiv& \ddot{h}_{\mu\nu},
\end{eqnarray}
the next step consists in redefining the action by means of Lagrange multipliers
\begin{eqnarray}
\nonumber
S' &=& \int\mathcal{L}dt + \int[\lambda_{1}^{\mu\nu}(\dot{h}_{\mu\nu}-s_{\mu\nu}) + \lambda_{2}^{\mu\nu}(\dot{s}_{\mu\nu}-v_{\mu\nu})]d^{4}x
\\
&\equiv& \int\mathcal{L}'dt,
\end{eqnarray}
where the order-reduced Lagrangian density $\mathcal{L}$ reads
\begin{widetext}
\begin{eqnarray}
\nonumber
\mathcal{L} &=& \int \Big[ f(0)\Big( 1 - \frac{1}{2}h_{00} + \frac{1}{2}h^{i}{}_{i} - \frac{1}{8}(h_{00})^{2} - \frac{1}{4}h_{00}h^{i}{}_{i} + \frac{1}{8}(h^{i}{}_{i})^{2} - \frac{1}{2}h^{0i}h_{0i} - \frac{1}{4}h^{ij}h_{ij} \Big)
\\
\nonumber
&& + f'(0)\Big( \frac{1}{2}s^{ij}s_{ij} - \frac{1}{2}s^{i}{}_{i}s^{j}{}_{j} + h^{0i}\nabla^{2}h_{0i} + \frac{1}{2}h^{ij}\nabla^{2}h_{ij} + 2s^{i}{}_{i}\partial^{j}h_{0j} - h_{00}\partial_{i}\partial_{j}h^{ij}
\\
\nonumber
&& + h^{k}{}_{k}\partial_{i}\partial_{j}h^{ij} - 2s^{ij}\partial_{i}\partial_{i}h_{0j} + h_{0i}\partial^{i}\partial^{j}h_{0j} - h_{i}{}^{k}\partial^{i}\partial_{j}h^{j}{}_{k} + h_{00}\nabla^{2}h^{i}{}_{i} - \frac{1}{2}h^{i}{}_{i}\nabla^{2}h^{j}{}_{j} \Big)
\\
&& + f''(0)\Big( \big[v^{i}{}_{i} - 2\partial^{i}s_{0i} + \partial^{i}\partial^{j}h_{ij} + \nabla^{2}h_{00} - \nabla^{2}h^{i}{}_{i}\big]\phi - \frac{1}{2}\phi^{2} \Big) \Big]d^{3}x.
\end{eqnarray}
\end{widetext}
The canonical momenta are calculated from $\mathcal{L}'$ and are given by:
\begin{eqnarray}
\label{CanonicalMomenta1}
\pi^{\mu\nu}&\equiv&\frac{\partial\mathcal{L}'}{\partial\dot{h}_{\mu\nu}} = \Lambda_{1}^{\mu\nu}, 
\\
\label{CanonicalMomenta2}
p^{\mu\nu}&\equiv&\frac{\partial\mathcal{L}'}{\partial\dot{s}_{\mu\nu}} = \Lambda_{2}^{\mu\nu},
\\
\label{CanonicalMomenta3}
p_{\phi}&\equiv&\frac{\partial\mathcal{L}'}{\partial\dot{\phi}} = 0,
\\
\label{CanonicalMomenta4}
p_{v}^{\mu\nu}&\equiv&\frac{\partial\mathcal{L}'}{\partial\dot{v}_{\mu\nu}} = 0,
\\
\label{CanonicalMomenta5}
\Lambda^{1}_{\mu\nu}&\equiv&\frac{\partial\mathcal{L}'}{\partial\dot{\lambda}_{1}^{\mu\nu}} = 0,
\\
\label{CanonicalMomenta6}
\Lambda^{2}_{\mu\nu}&\equiv&\frac{\partial\mathcal{L}'}{\partial\dot{\lambda}_{2}^{\mu\nu}} = 0,
\end{eqnarray}
and according to \cite{Gitman} the primary constraints are given by $\varphi^{\mu\nu}\equiv p^{\mu\nu}-\frac{\partial\mathcal{L}'}{\partial v_{\mu\nu}}\approx0$, and we also must not forget about $\varphi_{\phi}\equiv p_{\phi}\approx0$. Here $\approx$ means that the equality holds only in constraint surface in phase-space \cite{Henneaux}.

By definition, the canonical Hamiltonian density is given by $\mathcal{H}_{0}=\int(\pi^{\mu\nu}\dot{h}_{\mu\nu}+p^{\mu\nu}\dot{s}_{\mu\nu}+p_{v}^{\mu\nu}\dot{v}_{\mu\nu}+\varphi_{\phi}\dot{\phi}+\Lambda_{\mu\nu}^{1}\dot{\lambda}_{1}^{\mu\nu}+\Lambda_{\mu\nu}^{2}\dot{\lambda}_{2}^{\mu\nu})d^{3}x - \mathcal{L}'$, after replacing (\ref{CanonicalMomenta1})-(\ref{CanonicalMomenta6}) in $\mathcal{H}_{0}$, and thus, dropping out the parameters $(\lambda^{1}_{\mu\nu},\lambda^{2}_{\mu\nu})$ in favour of the canonical momenta $(\pi^{\mu\nu},p^{\mu\nu})$ we get
\begin{widetext}
\begin{eqnarray}
\nonumber
\mathcal{H}_{0} &=& \int\Big[ \pi^{\mu\nu}s_{\mu\nu} + p^{\mu\nu}v_{\mu\nu} - f(0)\Big( 1 - \frac{1}{2}h_{00} + \frac{1}{2}h^{i}{}_{i} - \frac{1}{8}(h_{00})^{2} - \frac{1}{4}h_{00}h^{i}{}_{i} + \frac{1}{8}(h^{i}{}_{i})^{2} - \frac{1}{2}h^{0i}h_{0i} - \frac{1}{4}h^{ij}h_{ij} \Big)
\\
\nonumber
&& - f'(0)\Big( \frac{1}{2}s^{ij}s_{ij} - \frac{1}{2}s^{i}{}_{i}s^{j}{}_{j} + h^{0i}\nabla^{2}h_{0i} + \frac{1}{2}h^{ij}\nabla^{2}h_{ij} + 2s^{i}{}_{i}\partial^{j}h_{0j} - h_{00}\partial_{i}\partial_{j}h^{ij} + h^{k}{}_{k}\partial_{i}\partial_{j}h^{ij}
\\
\nonumber
&& - 2s^{ij}\partial_{i}h_{0j} + h_{0i}\partial^{i}\partial^{j}h_{0j} - h_{i}{}^{k}\partial^{i}\partial_{j}h^{j}{}_{k} + h_{00}\nabla^{2}h^{j}{}_{j} - \frac{1}{2}h^{i}{}_{i}\nabla^{2}h^{k}{}_{k} \Big) - f''(0)\Big( \big[v^{i}{}_{i} - 2\partial^{i}s_{0i}
\\
\label{CanonicalHamiltonian}
&& + \partial^{i}\partial^{j}h_{ij} + \nabla^{2}h_{00} - \nabla^{2}h^{i}{}_{i}\big]\phi - \frac{1}{2}\phi^{2} \Big) \Big]d^{3}x.
\end{eqnarray}
\end{widetext}
The variables $\lambda_{1}^{\mu\nu}$, $\lambda_{2}^{\mu\nu}$, $v_{\mu\nu}$ (and their momenta) had been served their purpose for defining a Hamiltonian and for identifying the primary constraints, thus we can discard them as true canonical variables. The only canonical variables are $(h_{\mu\nu},\pi^{\mu\nu},s_{\mu\nu},p^{\mu\nu},\phi,p_{\phi})$, consequently, the fundamental Poisson brackets at stake are:
\begin{eqnarray}
\lbrace h_{\mu\nu},\pi^{\alpha\beta} \rbrace &=& \frac{1}{2}(\delta_{\mu}^{\alpha}\delta_{\nu}^{\beta} + \delta_{\mu}^{\beta}\delta_{\nu}^{\alpha})\delta^{3}(x-y),
\\
\lbrace s_{\mu\nu},p^{\alpha\beta} \rbrace &=& \frac{1}{2}(\delta_{\mu}^{\alpha}\delta_{\nu}^{\beta} + \delta_{\mu}^{\beta}\delta_{\nu}^{\alpha})\delta^{3}(x-y),
\\
\lbrace \phi,\varphi_{\phi} \rbrace &=& \delta^{3}(x-y).
\end{eqnarray}
Now, since we have primary constraints $(\varphi^{00},\varphi^{0i},\varphi^{ij},\varphi_{\phi})$ we need to include them in the dynamics by defining the primary Hamiltonian
\begin{equation}
\mathcal{H}_{1} = \mathcal{H}_{0} + \int\big(\Delta_{00}^{\varphi}\varphi^{00} + \Delta_{0i}^{\varphi}\varphi^{0i} + \Delta_{ij}^{\varphi}\varphi^{ij} + \Delta^{\phi} \big)d^{3}x
\end{equation}
where the $\Delta's$ multipliers are enforcing the primary constraints given explicitly by:
\begin{eqnarray}
\varphi^{00} &\equiv& p^{00} \approx 0,
\\
\varphi^{0i} &\equiv& p^{0i} \approx 0,
\\
\varphi^{ij} &\equiv& p^{ij} - f''(0)\eta^{ij}\phi \approx 0,
\\
\varphi_{\phi} &=& p _{\phi} \approx 0.
\end{eqnarray}
Consistency of primary constraints $\dot{\varphi}\approx0$ lead us to new constraints
\begin{eqnarray}
\dot{\varphi}^{00}\approx 0 \quad &\Rightarrow& \quad \chi^{00} \equiv \pi^{00} \approx 0,
\\
\dot{\varphi}^{0i} \approx 0 \quad &\Rightarrow& \quad \chi^{0i} \equiv \pi^{0i} - f''(0)\partial^{i}\phi \approx 0,
\end{eqnarray}
and to relations for the $\Delta's$ multipliers 
\begin{eqnarray}
\nonumber
\dot{\varphi}^{ij} \approx 0 \quad &\Rightarrow& \quad \pi^{ij} - f'(0)\big(s^{ij} - \eta^{ij}s^{k}{}_{k} + 2\eta^{ij}\partial^{k}h_{0k}
\\
&& - \partial^{i}h_{0}{}^{j} - \partial^{j}h_{0}{}^{i} \big) + f''(0)\eta^{ij}\Delta^{\phi} \approx 0,
\\
\nonumber
\dot{\varphi}_{\phi} \approx 0 \quad &\Rightarrow& \quad f''(0)\big[ \phi - (v^{i}{}_{i} - 2\partial^{i}s_{0i} + \partial^{i}\partial^{j}h_{ij}
\\
&& + \nabla^{2}h_{00} - \nabla^{2}h^{i}{}_{i}) - \eta^{ij}\Delta_{ij}^{\varphi} \big] \approx 0.
\end{eqnarray}
Since we have new constraints it is required to apply consistency again for those new constraints, after repeating this algorithm \cite{Dirac1} until no more constraints appear and by manipulating the relations for the $\Delta's$ multipliers (more details in Appendix section) we arrive to the following set of constraints:
\begin{eqnarray}
\label{Constraints1}
\varphi^{00} &=& p^{00},
\\
\label{Constraints2}
\varphi^{0i} &=& p^{0i},
\\
\label{Constraints3}
\varphi^{ij} &=& p^{ij} - f''(0)\eta^{ij}\phi,
\\
\label{Constraints4}
\varphi_{\phi} &=& p_{\phi},
\\
\label{Constraints5}
\chi^{00} &=& \pi^{00},
\\
\label{Constraints6}
\chi^{0i} &=& \pi^{0i} - f''(0)\partial^{i}\phi,
\\
\nonumber
\gamma^{00} &=& f(0)(\frac{1}{2} + \frac{1}{4}h_{00} + \frac{1}{4}h^{i}{}_{i}) + f'(0)(\partial_{i}\partial_{j}h^{ij}
\\
\label{Constraints7}
&& - \nabla^{2}h^{i}{}_{i}) - f''(0)\nabla^{2}\phi,
\\
\label{Constraints8}
\psi^{0i} &=& \partial_{j}\pi^{ij} + \frac{1}{2}f(0)h_{0}{}^{i},
\\
\nonumber
\chi^{ij} &=& \pi^{ij} - f'(0)\big(s^{ij} - \eta^{ij}s^{k}{}_{k} + 2\eta^{ij}\partial^{k}h_{0k}
\\
\nonumber
&& - \partial^{i}h_{0}{}^{j} - \partial^{j}h_{0}{}^{i} \big) - \frac{1}{3}\eta^{ij}\big[ \pi^{k}{}_{k}
\\
\label{Constraints9}
&& + f'(0)(2s^{k}{}_{k} + 4\partial^{k}h_{0k}) \big],
\\
\label{Constraints9.1}
\tau &=& f(0)\big[ \partial^{i}h_{0i} - \frac{1}{2}(s_{00}+s^{i}{}_{i}) \big],
\\
\label{Constraints10}
\xi^{i} &=& f(0)\big[\frac{1}{2}(\partial^{i}h^{j}{}_{j}-\partial^{i}h_{00}) - \partial_{j}h^{ij} + s_{0}{}^{i} \big].
\end{eqnarray}
The constraints (\ref{Constraints1})-(\ref{Constraints10}) are quite general in the sense that no assumptions on $f(0)$, $f'(0)$ and $f''(0)$ are made but, now we must take $f(0)=0$ because of (\ref{Condition}) (remember that $R_{0}=0$). From now, we will separate the analysis in different cases depending upon conditions on $f'(0)$ and $f''(0)$.

\subsection{Case 1: $f'(0)\neq0$ and $f''(0)\neq0$}

This sub-section is devoted to the $(R_{0}=0)-$\textit{non-degenerate} case. The condition $f(0)=0$ implies that the constraints (\ref{Constraints9.1}) and (\ref{Constraints10}) disappears, thus, the only constraints at stake are (\ref{Constraints1})-(\ref{Constraints9}) without the terms proportional to $f(0)$. In order to count the number of degrees of freedom we need to classify the constraints, for this purpose we can construct the matrix $C^{IJ}$ whose entries are the Poisson brackets between constraints, such matrix acquires the following form
\begin{widetext}
\begin{equation}
C^{IJ} =
\bordermatrix{
 & \varphi^{00} & \varphi^{0k} & \varphi^{kl} & \varphi_{\phi} & \chi^{00} & \chi^{0k} & \gamma^{00} & \psi^{0k} & \chi^{kl} & \cr
\varphi^{00} & 0 & 0 & 0 & 0 & 0 & 0 & 0 & 0 & 0 & \cr
\varphi^{0i} & 0 & 0 & 0 & 0 & 0 & 0 & 0 & 0 & 0 & \cr
\varphi^{ij} & 0 & 0 & 0 & \lbrace \varphi^{ij},\varphi_{\phi} \rbrace & 0 & 0 & 0 & 0 & \lbrace \varphi^{ij},\chi^{kl} \rbrace & \cr
\varphi_{\phi} & 0 & 0 & \lbrace \varphi_{\phi},\varphi^{kl} \rbrace & 0 & 0 & \lbrace \varphi_{\phi},\chi^{0k} \rbrace & \lbrace \varphi_{\phi},\gamma^{00} \rbrace & 0 & 0 & \cr
\chi^{00} & 0 & 0 & 0 & 0 & 0 & 0 & 0 & 0 & 0 & \cr
\chi^{0i} & 0 & 0 & 0 & \lbrace \chi^{0i},\varphi_{\phi} \rbrace & 0 & 0 & 0 & 0 & \lbrace \chi^{0i},\chi^{kl} \rbrace & \cr
\gamma^{00} & 0 & 0 & 0 & \lbrace \gamma^{00},\varphi_{\phi} \rbrace & 0 & 0 & 0 & 0 & \lbrace \gamma_{00},\chi^{kl} \rbrace & \cr
\psi^{0i} & 0 & 0 & 0 & 0 & 0 & 0 & 0 & 0 & 0 & \cr
\chi^{ij} & 0 & 0 & \lbrace \chi^{ij},\varphi^{kl} \rbrace & 0 & 0 & \lbrace \chi^{ij},\chi^{0k} \rbrace & \lbrace \chi^{ij},\gamma_{00} \rbrace & 0 & 0 & \cr
}
\end{equation}
\end{widetext}
where the indicated Poisson brackets are:
\begin{eqnarray}
\nonumber
\lbrace \varphi^{ij},\varphi_{\phi} \rbrace &=& - f''(0)\eta^{ij}\delta^{3}(x-y),
\\
\nonumber
\lbrace \varphi^{ij},\chi^{kl} \rbrace &=& f'(0)\big[\frac{1}{2}(\eta^{ik}\eta^{jl}+\eta^{il}\eta^{jk}) - \eta^{ij}\eta^{kl} \big]\delta^{3}(x-y),
\\
\nonumber
\lbrace \varphi_{\phi},\chi^{0k} \rbrace &=& f''(0)\partial^{k}\delta^{3}(x-y),
\\
\nonumber
\lbrace \varphi_{\phi},\gamma^{00} \rbrace &=& f''(0)\nabla^{2}\delta^{3}(x-y),
\\
\nonumber
\lbrace \chi^{0i},\chi^{kl} \rbrace &=& f'(0)\big[\eta^{kl}\partial^{i} - \frac{1}{2}(\eta^{il}\partial^{k}+\eta^{ik}\partial^{l}) \big]\delta^{3}(x-y),
\\
\lbrace \gamma^{00},\chi^{kl} \rbrace &=& f'(0)(\partial^{k}\partial^{l}-\eta^{kl}\nabla^{2})\delta^{3}(x-y).
\end{eqnarray}
The null vectors of $C^{IJ}$ imply a reconfiguration of the constraints, in particular, the contraction of the constraints $(\varphi^{00},\varphi^{0i},\varphi^{ij},\varphi_{\phi},\chi^{00},\chi^{0i},\gamma^{00},\psi^{0i},\chi^{ij})$ with the null vectors $(0,0,-\partial_{i},0,0,1,0,0,0)$ and $(0,0,-\partial_{i}\partial_{j},0,0,0,1,0,0)$ give rise to the first-class constraints $\pi^{0i} - \partial_{j}p^{ij}$ and $f'(0)(\partial_{i}\partial_{j}h^{ij} - \nabla^{2}h^{i}{}_{i}) - \partial_{i}\partial_{j}p^{ij}$ respectively. Here we recall that first-class constraints are those whose Poisson bracket with all constraints vanish, otherwise, there are second-class. The full set of constraints is:
\begin{eqnarray}
\label{AConstraints1}
\Gamma_{1} &=& p^{00},
\\
\label{AConstraints2}
\Gamma_{2} &=& \pi^{00},
\\
\label{AConstraints3}
\Gamma_{3}^{i} &=& \pi^{0i} - \partial_{j}p^{ij},
\\
\label{AConstraints4}
\Gamma_{4}^{i} &=& \partial_{j}\pi^{ij},
\\
\label{AConstraints5}
\Gamma_{5} &=& f'(0)(\partial_{i}\partial_{j}h^{ij} - \nabla^{2}h^{i}{}_{i}) - \partial_{i}\partial_{j}p^{ij},
\\
\label{AConstraints6}
\Gamma_{6}^{i} &=& p^{0i},
\\
\label{AConstraints7}
\Omega_{1} &=& p_{\phi},
\\
\label{AConstraints8}
\Omega_{2}^{ij} &=& p^{ij} - f''(0)\eta^{ij}\phi,
\\
\nonumber
\Omega_{3}^{ij} &=& \pi^{ij} - f'(0)[s^{ij}-\eta^{ij}s^{k}{}_{k} + 2\eta^{ij}\partial^{k}h_{0k}
\\
\nonumber
&& - \partial^{i}h_{0}{}^{j} - \partial^{j}h_{0}{}^{i}] - \frac{1}{3}\eta^{ij}\big[\pi^{k}{}_{k}
\\
\label{AConstraints9}
&& + f'(0)(2s^{k}{}_{k}-4\partial^{k}h_{0k})\big].
\end{eqnarray}
The constraints $\Gamma's$ are first-class and the constraints $\Omega's$ are second-class. Now we are ready to proceed with the counting of degrees of freedom: we have 20 canonical variables from the pair $(h_{\mu\nu},\pi^{\mu\nu})$, the variables $(p_{\mu\nu},s^{\mu\nu})$ counts for another 20, the pair $(\phi,p_{\phi})$ adds 2 more; the number of first-class constraints is 12, as for the second-class constraints we should be careful in the counting, that is, the constraint $\Omega_{3}^{ij}$ only counts as 5 independent constraints because it is traceless, the total number of second-class constraints is 12. Thus we have
\begin{equation}
DoF = \frac{1}{2}\big[42 - 2(12) - 12\big] = 3.
\end{equation}
This results confirms that an $f(R)$ model of gravity propagates 3 degrees of freedom provided that $f(R)$ is $(R_{0}=0)-$\textit{non-degenerate} and only if $f''(0)\neq0$. This is a well-known result for \textit{Einstein}$-R^{2}$ gravity, i.e. when the function $f(R)$ have the form $f(R)=R+\alpha R^{2}$, $\alpha\neq0$ \cite{Stelle2,Starobinsky}.

\subsection{Case 2: $f'(R_{0})=0$ and $f''(R_{0})\neq0$}

Here we are entering in the $(R_{0}=0)-$\textit{degenerate} case. Now in addition to $f(0)=0$, we also need to make $f'(0)=0$ effective in the constraints (\ref{Constraints1})-(\ref{Constraints9}), on the other hand, in this case the relations for the $\Delta's$ multipliers lead us to identify the extra constraint $\nabla^{2}\pi^{i}{}_{i}\approx0$. The nullity of the reformulated $C^{IJ}$ matrix renders the following constraints configuration:
\begin{eqnarray}
\label{BConstraint1}
\Gamma_{1} &=& p^{00},
\\
\label{BConstraint2}
\Gamma_{2} &=& \pi^{00},
\\
\label{BConstraint3}
\Gamma_{3}^{i} &=& \pi^{0i} - \partial_{i}p^{ij},
\\
\label{BConstraint4}
\Gamma_{4}^{i} &=& \partial_{j}\pi^{ij},
\\
\label{BConstraint5}
\Gamma_{5}^{i} &=& p^{0i},
\\
\label{BConstraint6}
\Gamma_{6}^{ij} &=& \pi^{ij} - \frac{1}{3}\eta^{ij}\pi^{k}{}_{k},
\\
\label{BConstraint7}
\Gamma_{7} &=& \nabla^{2}\pi^{i}{}_{i},
\\
\label{BConstraint8}
\Omega_{1}^{ij} &=& p^{ij} - f''(0)\eta^{ij}\phi,
\\
\label{BConstraint9}
\Omega_{2} &=& p_{\phi},
\\
\label{BConstraint10}
\Omega_{3} &=& - f''(0)\nabla^{2}\phi.
\end{eqnarray}
Is important to note that now the constraint $\pi^{ij}-\frac{1}{3}\eta^{ij}\pi^{k}{}_{k}$ becomes first-class and that the constraint proportional to $\partial_{i}\partial_{j}h^{ij}-\nabla^{2}h^{i}{}_{i}$ have its reminiscence in $\Omega_{3}$ but now, as a second-class constraint. Since we have the classification of constraints at hand, we can proceed with the counting of degrees of freedom: we have 42 canonical variables as before, the number of first-class constraints is 17 (we discard again one of the $\Gamma_{6}^{ij}$ due to its lack of trace), and the number of second-class constraints is 8, so we have that
\begin{equation}
DoF = \frac{1}{2}\big[42 - 2(17) - 8\big] = 0.
\end{equation}
In this case the theory lacks of propagating degrees of freedom, the apparent scalaron mode is not present, that is, the Minkowski spacetime is strongly-coupled to the $(R_{0}=0)-$\textit{degenerate} $f(R)$ models. In particular, this result corroborates the findings of recent works for \textit{Pure}$-R^{2}$ gravity, $f(R)=\alpha R^{2}$ \cite{Hell,Barker}.

\subsection{Case 3: $f'(0)\neq0$ and $f''(0)=0$}

Now we are back to the $(R_{0}=0)-$\textit{non-degenerate} case but, with the distinction that now $f''(0)$ also vanish. We turn once again to the constraints (\ref{Constraints1})-(\ref{Constraints9}) and eliminate any contribution of $f''(0)$ and, in contrast to the previous case, the relations for the $\Delta's$ multipliers gives no new information. By the same \textit{modus operandi} as in previous cases, that is, by null vector inspection we find the following constraints:
\begin{eqnarray}
\label{CConstraints1}
\Gamma_{1} &=& p^{00},
\\
\label{CConstraints2}
\Gamma_{2} &=& \pi^{00},
\\
\label{CConstraints3}
\Gamma_{3}^{i} &=& \pi^{0i} - \partial_{j}p^{ij},
\\
\label{CConstraints4}
\Gamma_{4}^{i} &=& \partial_{j}\pi^{ij},
\\
\label{CConstraints5}
\Gamma_{5} &=& f'(0)(\partial_{i}\partial_{j}h^{ij} - \nabla^{2}h^{i}{}_{i}) - \partial_{i}\partial_{j}p^{ij},
\\
\label{CConstraints6}
\Gamma_{6}^{i} &=& p^{0i},
\\
\label{CConstraints7}
\Gamma_{7} &=& p_{\phi},
\\
\label{CConstraints8}
\Omega_{1}^{ij} &=& p^{ij},
\\
\nonumber
\Omega_{2}^{ij} &=& \pi^{ij} - f'(0)\big(s^{ij} - \eta^{ij}s^{k}{}_{k} + 2\eta^{ij}\partial^{k}h_{0k}
\\
\label{CConstraints9}
&& - \partial^{i}h_{0}{}^{j}).
\end{eqnarray}
The counting is straightforward: we have 42 canonical variables, 13 first-class constraints and 12 second-class constraints because we cannot conclude from (\ref{CConstraints9}) that the constraint $\Omega_{2}^{ij}$ is traceless, the resulting number is
\begin{equation}
DoF = \frac{1}{2}\big[42 - 2(13) - 12\big] = 2.
\end{equation}
Alternatively, we can argue that the action loses its higher-order character and then, we can evade the higher-order treatment, in this approach the only surviving constraints are:
\begin{eqnarray}
\label{C2Constraints1}
\Gamma_{1} &=& \pi^{00},
\\
\label{C2Constraints2}
\Gamma_{2}^{i} &=& \pi^{0i},
\\
\label{C2Constraints3}
\Gamma_{3}^{i} &=& \partial_{j}\pi^{ij},
\\
\label{C2Constraints4}
\Gamma_{4} &=& f'(0)(\partial^{i}\partial^{j}h_{ij} - \nabla^{2}h^{i}{}_{i}),
\end{eqnarray}
with canonical Hamiltonian density given by
\begin{eqnarray}
\nonumber
\mathcal{H}_{0} &=& \int\Big[ \frac{1}{2f'(0)}(\pi^{ij}\pi_{ij}-\frac{1}{2}\pi^{i}{}_{i}\pi^{j}{}_{j}) - 2\partial_{i}\pi^{ij}h_{0j}
\\
\nonumber
&& - f'(0)\Big( \frac{1}{2}h^{ij}\nabla^{2}h_{ij} - h_{00}\partial_{i}\partial_{j}h^{ij}
\\
\nonumber
&& + h^{k}{}_{k}\partial_{i}\partial_{j}h^{ij} - h_{i}{}^{k}\partial^{i}\partial_{j}h^{j}{}_{k} + h_{00}\nabla^{2}h^{j}{}_{j}
\\
&& - \frac{1}{2}h^{i}{}_{i}\partial^{j}\partial_{j}h^{k}{}_{k} \Big) \Big]d^{4}x,
\end{eqnarray}
such Hamiltonian is well-defined because we are assuming that $f'(0)\neq0$, the number of degrees of freedom retains the same, as expected
\begin{equation}
DoF = \frac{1}{2}\big[20 - 2(8) - 0\big] = 2.
\end{equation}
We can conclude that any $(R_{0}=0)-$\textit{non-degenerate} $f(R)$ model propagates 2 degrees of freedom, provided that $f''(0)=0$, this number is somewhat anticipated because the scalaron mode is not present. The result is also pertinent if we recall standard linearized gravity, $f(R)=R$ \cite{Barcelos}.

\subsection{Case 4: $f'(0)=0$ and $f''(0)=0$}

To conclude the analysis, we consider the $(R_{0}=0)-$\textit{degenerate} case when $f''(0)=0$. It is clear that if we take $f(0)=0$, $f'(0)=0$ and $f''(0)=0$ then the action (\ref{Action}) is identically zero, thus no propagating degrees of freedom are present. If we insist in the constraints (\ref{Constraints1})-(\ref{Constraints9}) we can do a naive counting by considering the following set:
\begin{eqnarray}
\Gamma_{1} &=& p^{00},
\\
\Gamma_{2} &=& \pi^{00},
\\
\Gamma_{3}^{i} &=& \pi^{0i},
\\
\Gamma_{4}^{i} &=& p^{0i},
\\
\Gamma_{5} &=& p_{\phi},
\\
\Gamma_{6}^{ij} &=& p^{ij},
\\
\Gamma_{7}^{ij} &=& \pi^{ij},
\end{eqnarray}
with canonical Hamiltonian density given by
\begin{equation}
\mathcal{H}_{0} = \int\big( \pi^{\mu\nu}s_{\mu\nu} + p^{\mu\nu}v_{\mu\nu} \big) d^{3}x.
\end{equation}
With all the 21 first-class constraints at hand we reiterate the number of degrees of freedom (again, no traceless implication on $\Gamma_{7}^{ij}$):
\begin{eqnarray*}
DoF &=& \frac{1}{2}\big[42 - 2(21) - 0\big] = 0.
\end{eqnarray*}
One way or another, we found that the spectrum of the theory is empty. As example, we can think of any theory given by $f(R)=\alpha R^{k}$, where $k>2$ and $\alpha\neq0$.

\section{Conclusions}

In this work, we developed the Hamiltonian formulation of the $(R_{0}=0)-$\textit{degenerate} and $(R_{0}=0)-$\textit{non-degenerate} $f(R)$ gravity models around Minkowski background, we identified all the constraints of the theory by following the appropriate methods. As expected, some of the constraints (and some of the relations for Lagrange multipliers) had its dependence on $f(0)$, $f'(0)$ or $f''(0)$, in particular $f''(0)$ is always present in the constraints associated to the scalar field $\phi$, this is because $\phi$ is directly attached to the second-order term of the expansion (\ref{Expansion}) and therefore, to the second derivative $f''(0)$. In the general set (\ref{Constraints1})-(\ref{Constraints10}), we made $f(0)=0$ effective and then we separated the study in four cases, two cases for the $(R_{0}=0)-$\textit{non-degenerate} models and another two for the $(R_{0}=0)-$\textit{degenerate} models, depending whether $f''(0)=0$ or $f''(0)\neq0$. This division was convenient for handling the constraints of the theory and to make a transparent counting of the degrees of freedom, lets make some comments on each particular case.

First, as for the less restrictive case that comes from assuming $f'(0)\neq0$ and $f''(0)\neq0$, we found that these $f(R)$ models propagate 3 degrees of freedom, associated with two massless spin-2 modes and to the scalaron (see equation (\ref{KG})). It is pertinent to mention that the count made in \cite{Liang} also recognizes 3 degrees of freedom, this is achieved by invoking the equivalence of $f(R)$ gravity with scalar-tensor theory (that is, by assuming $f'(R)>0$ and $f''(R)\neq0$) \cite{O'Hanlon}, no comment is made on the discontinuity in the degrees of freedom. Same conclusion is obtained in \cite{Myung} and this agreement takes place because the model studied in \cite{Myung} is such that $f(0)=0$, $f'(0)=1$ and $f''(0)=\frac{1}{3M^{2}}$ being $M>0$ a mass parameter, however, we differ from their final statement, namely: \textit{DoF counting of $f(R)$ gravity theories should be independent of propagating space-time}, instead we may state: \textit{in performing a perturbation theory, DoF counting of $f(R)$ gravity theories could be dependent of propagating space-time}, as we shall see immediately.

As for the second case ($f'(0)=0$ and $f''(0)\neq0$), the constraints were reconfigured in the following manner: the relations for the multipliers allowed us to identify an additional constraint (\ref{BConstraint7}), the constraint $\pi^{ij}-\frac{1}{3}\eta^{ij}\pi^{k}{}_{k}$ was promoted to first-class and the constraint initially proportional to $\partial_{i}\partial_{j}h^{ij}-\nabla^{2}h^{i}{}_{i}$ became second-class, such reconfiguration annihilated all the degrees of freedom, therefore, no massless scalar mode was present. Thus, the Hamiltonian formalism confirmed us that Minkowski spacetime can be strongly-coupled to $f(R)$ gravity theories, that is, some of the degrees of freedom (all of them, in this case) does not propagate on top of it. We recall that this seems to be an artefact of perturbation theory, manifestly in the choice of the background.

Regarding the third case ($f'(0)\neq0$ and $f''(0)=0$), the constraint $p_{\phi}$ became first-class because the field $\phi$ was not present any more in the constraints, on the other hand, the relations for multipliers does not implied new constraints, the subsequent counting gave us 2 propagating degrees of freedom, additionally, the same number was recovered by bypassing the higher-order treatment. The story repeats in the sense that not all $f(R)$ models can propagate the 3 well-known modes, certainly, having only 2 degrees of freedom is not unnatural after all, in other words, we can think of this case as the most similar to standard linearized gravity for many reasons: the presence of the kinetic term in the action, the form of the canonical Hamiltonian and the structure of the constraints, so it is totally reasonable to obtain 2 propagating degrees of freedom even thought $f'(0)$ is really unspecified.

Finally, in the case where $f'(0)=0$ and $f''(0)=0$ we found no propagating degrees of freedom simply because $S$ vanished identically, alternatively, by clinging to the constraints we were able to reaffirm the empty spectrum. Based on this result we can assert that, if one hopes to perform perturbation theory around the Minkowski spacetime on an $f(R)$ model, then, any polynomial expression $f(R)=\alpha_{k}R^{k}$ with $k>2$ will not constitute a physically viable option.

In the following table, we summarize the results that our Hamiltonian analysis was able to reproduce, as well as basic examples for each case
\begin{table}[ht]
\renewcommand{\arraystretch}{1.2}
\caption{Spectrum of $f(R)$ gravity around Minkowski spacetime.}
\begin{ruledtabular}
\begin{tabular}{ccc}
Condition & Degrees of freedom & Example \\
\hline
$f'(0)\neq0$, $f''(0)\neq0$ & 3 & $f(R)=R + \alpha R^{2}$ \\
$f'(0)\neq0$, $f''(0)=0$ & 2 & $f(R)=R$ \\
$f'(0)=0$, $f''(0)\neq0$ & 0 & $f(R)=\alpha R^{2}$ \\
$f'(0)=0$, $f''(0)=0$ & 0 & $f(R)=\alpha R^{k}$, $k>2$ \\
\end{tabular}
\end{ruledtabular}
\end{table}

As an additional comment, we can point out that the canonical Hamiltonian (\ref{CanonicalHamiltonian}) exhibits an apparent instability (an Ostrogradsky type of instability \cite{Ostrogradsky}) because of the presence of the linear terms in the momenta $\pi^{\mu\nu}$ and $p^{\mu\nu}$ this could represent a problem but, it is known that those instabilities can be removed by eliminating the second-class constraints by means of the Dirac brackets \cite{Ganz}, see for example \cite{Escalante} where the canonical Hamiltonian was successfully \textit{healed} by this procedure. Considering that we have properly identified the second-class constraints in each case, then the Dirac brackets can be calculated, since this goes beyond the main scope of this work, we will leave this issue for another instance.

In the end, the results of this work confirm some of the outcomes of the recent literature \cite{Casado1,Casado2}, in particular, that Minkowski spacetime can be strongly-coupled to some $f(R)$ models, i.e., that the usual 3 degrees of freedom will not always be present. Since these conclusions were derived using different methods from those presented in \cite{Casado1,Casado2,Liang,Stelle2,Starobinsky,Myung}, we believe that this work offers a contribution to the current discussion around $f(R)$ gravity.

\begin{acknowledgments}
The author would like to thank A. Escalante for discussion of the topic and reading of the manuscript.
\end{acknowledgments}

\appendix*
\section{}

For the sake of completeness, in this section we clarify some intermediate steps in the identification process of the constraints (\ref{Constraints1})-(\ref{Constraints10}). Consistency of $\chi^{00}$ and $\chi^{0i}$ gives the constraint $\gamma^{00}$ and the relation (\ref{DeltaMultipliers3}) respectively, then from consistency of $\gamma^{00}$ we obtain the relation (\ref{DeltaMultipliers4}). We can immediately note that consistency condition gives no new constraints but, this fact does not mean that we cannot identify further constraints by another means, at this point we have at hand the constraints (\ref{Constraints1})-(\ref{Constraints7}) and the following relations for the multipliers:
\begin{widetext}
\begin{eqnarray}
\label{DeltaMultipliers1}
\pi^{ij} - f'(0)\big(s^{ij} - \eta^{ij}s^{k}{}_{k} + 2\eta^{ij}\partial^{k}h_{0k} - \partial^{i}h_{0}{}^{j} - \partial^{j}h_{0}{}^{i} \big) + f''(0)\eta^{ij}\Delta^{\phi} &\approx& 0,
\\
\label{DeltaMultipliers2}
f''(0)\big[ \phi - (v^{i}{}_{i} - 2\partial^{i}s_{0i} + \partial^{i}\partial^{j}h_{ij} + \nabla^{2}h_{00} - \nabla^{2}h^{i}{}_{i}) - \eta^{ij}\Delta_{ij}^{\varphi} \big] &\approx& 0,
\\
\label{DeltaMultipliers3}
\frac{1}{2}f(0)h_{0}{}^{i} + f'(0)(\partial^{i}\partial^{j}h_{0j}-\nabla^{2}h_{0}{}^{i} + \partial_{j}s^{ij} - \partial^{i}s^{j}{}_{j}) - f''(0)\partial^{i}\Delta^{\phi} &\approx& 0,
\\
\label{DeltaMultipliers4}
\frac{1}{4}f(0)(s_{00} + s^{i}{}_{i}) + f'(0)(\partial^{i}\partial^{j}s_{ij}-\eta^{ij}\nabla^{2}s_{ij}) - f''(0)\nabla^{2}\Delta^{\phi} &\approx& 0.
\end{eqnarray}
\end{widetext}
These relations reveals more constraints. First, if we take the derivative $\partial_{i}$ of (\ref{DeltaMultipliers1}) and then combine the result with relation (\ref{DeltaMultipliers3}) then we identify the constraint (\ref{Constraints8}). Second, if we contract (\ref{DeltaMultipliers1}) with $\eta_{ij}$ and then we subtract the result to the same relation (\ref{DeltaMultipliers1}) then we obtain the constraint (\ref{Constraints9}). On the other hand, we can take (\ref{DeltaMultipliers4}) and subtract it to $\partial_{i}$(\ref{DeltaMultipliers3}), this procedure gives the constraint (\ref{Constraints9.1}). Dirac's algorithm demands to also apply consistency on these constraints, evolution of $\psi^{0i}$, $\chi^{ij}$ and $\tau$ leads to the constraint (\ref{Constraints10}) and to the following relations respectively:
\begin{widetext}
\begin{eqnarray}
\nonumber
&& f(0)\Big( \frac{1}{2}\eta^{ij} - \frac{1}{4}h_{00}\eta^{ij} + \frac{1}{4}h^{k}{}_{k}\eta^{ij} - \frac{1}{2}h^{ij} + \big[\frac{1}{4}\partial^{k}s_{0k} - \frac{1}{8}(v_{00}+v^{k}{}_{k})\big]\eta^{ij} \Big) + f'(0)\Big( \nabla^{2}h^{ij} 
\\
\nonumber
&& - \partial^{i}\partial^{j}h_{00} + \partial^{i}\partial^{j}h^{k}{}_{k} + [\partial^{k}\partial^{l}h_{kl} + \nabla^{2}h_{00} - \nabla^{2}h^{k}{}_{k}]\eta^{ij} - [\partial^{i}\partial^{k}h^{j}{}_{k}+\partial^{j}\partial^{k}h^{i}{}_{k}] - 2\eta^{ij}\partial^{k}s_{0k} 
\\
\nonumber
&& + [\partial^{i}s_{0}{}^{j}+\partial^{i}s_{0}{}^{j}] - v^{ij} + v^{k}{}_{k}\eta^{ij} \Big) + f''(0)\big( \partial^{i}\partial^{j}\phi - \eta^{ij}\nabla^{2}\phi \big) + \frac{1}{2}f'(0)\big[\eta^{ij}\eta^{kl} - (\eta^{ik}\eta^{jl}
\\
\nonumber
&& +\eta^{il}\eta^{jk}) \big]\Delta_{kl}^{\varphi} - \frac{1}{3}\Big[ f(0)\Big( \frac{3}{2} - \frac{3}{4}h_{00} + \frac{1}{4}h^{i}{}_{i}\big[\frac{1}{4}\partial^{i}s_{0i} - \frac{1}{8}(v_{00}+v^{i}{}_{i})\big]\Big) + f'(0)\Big( - \nabla^{2}h^{i}{}_{i}
\\
\label{DeltaMultipliers5}
&& + 2\nabla^{2}h_{00} + \partial^{i}\partial^{j}h_{ij} - 4\partial^{i}s_{0i} + 2v^{i}{}_{i}\Big) - 3f''(0)\nabla^{2}\phi + \frac{1}{2}f'(R_{0})\eta^{ij}\Delta_{ij}^{\varphi} \Big] \approx 0,
\label{DeltaMultipliers6}
\\
&& f(0)(\partial^{i}s_{0i} - v_{00} - v^{i}{}_{i}) - \Delta_{00}^{\varphi} - \eta^{ij}\Delta_{ij}^{\varphi} \approx 0.
\end{eqnarray}
\end{widetext}
Finally, consistency of (\ref{Constraints10}) only lead us to the following relation
\begin{equation}
\label{DeltaMultipliers7}
2v_{0}{}^{i} - \partial^{i}s_{00} + \partial^{i}s^{j}{}_{j} - 2\partial_{j}s^{ij} + \eta^{ij}\Delta_{0j}^{\varphi} \approx 0,
\end{equation}
thus, we reach the end of the identification process because no more constraints arise from Dirac's algorithm and because the relations (\ref{DeltaMultipliers1})-(\ref{DeltaMultipliers7}) do not provide new information.

\bibliography{References}

\end{document}